# Uncovering the dynamic precursors to motor-driven contraction of active gels

José Alvarado[1,2,$], Luca Cipelletti[3,*], Gijsje Koenderink[1,*]

[1] AMOLF, Physics of Living Matter Department, 1098 XG Amsterdam, Netherlands
[2] Massachusetts Institute of Technology, Department of Mechanical Engineering, Cambridge, MA, 02139, USA
[3] L2C, Univ. Montpellier, CNRS, Montpellier, France
[$] Current address: University of Texas at Austin, Department of Physics, Austin, TX, 78712, USA
[*] Corresponding authors. LC: luca.cipelletti@umontpellier.fr; GK: g.koenderink@amolf.nl

## Abstract

Cells and tissues have the remarkable ability to actively generate the forces required to change their shape. This active mechanical behavior is largely mediated by the actin cytoskeleton, a crosslinked network of actin filaments that is contracted by myosin motors. Experiments and active gel theories have established that the length scale over which gel contraction occurs is governed by a balance between molecular motor activity and crosslink density. By contrast, the dynamics that govern the contractile activity of the cytoskeleton remain poorly understood. Here we investigate the microscopic dynamics of reconstituted actin-myosin networks using simultaneous real-space video microscopy and Fourier-space dynamic light scattering. Light scattering reveals rich and unanticipated microscopic dynamics that evolve with sample age. We uncover two dynamical precursors that precede macroscopic gel contraction. One is characterized by a progressive acceleration of stress-induced rearrangements, while the other consists of sudden rearrangements that depend on network adhesion to the boundaries and are highly heterogeneous. Our findings reveal an intriguing analogy between self-driven rupture and collapse of active gels to the delayed rupture of passive gels under external loads.

## 1 Introduction

*Active matter* designates a fast-growing research area in soft condensed matter dealing with systems comprised of self-propelled constituents, as opposed to passive materials, for which thermal energy is the only driver of dynamics. Active soft systems are found in living organisms and there is also a growing class of synthetic variants that are for instance based on self-propelled colloids, molecular motors, or DNA [1].

*Active gels* are a class of active soft systems particularly relevant to biology. Their prototypical example is the cell cytoskeleton: a network of protein filaments that spans the cytoplasm and actively deforms the cell boundaries via pushing and pulling forces [2].

The most important contributor to cell shape changes is the actin cytoskeleton, which generates contractile forces with the help of myosin-II motors [3,4]. Myosin-II is a double-headed motor protein that organizes into bipolar filaments, which harness chemical energy derived from ATP hydrolysis to pull pairs of antiparallel actin filaments towards one another [5]. Although myosin filaments are only micrometer-sized and exert only pN-level forces, cells spatially integrate the activity of many motors to produce larger contractile forces (∼nN and greater) on cellular length scales [6]. The most efficient example of this sort of force integration occurs in striated muscle cells, where contraction occurs on force and length scales that approach the dimensions of the host organism. These contractions are mediated by a sarcomeric organization of the actin and myosin filaments, whose highly regular architecture promotes efficient force transmission [7]. Non-muscle cells lack such an ordered sarcomeric organization. They have a much more dynamic and adaptive actin-myosin cytoskeleton than muscle cells, which allows them to generate contractile forces on varying length scales. Myosin motors can exert localized pulling forces at the equator of dividing cells [8] and in the rear of migrating cells [9], but they can also generate cellular-length-scale pulling forces that facilitate cortical polarizing flows in developing oocytes [10]. Collectively, cells can even generate contractile forces that drive shape changes of entire tissues, which is important for embryonic development [11] and wound healing [12].

Until now most research on the physical basis of actomyosin contractility has focused on the question of how cells control the length scale on which contraction occurs. Single myosin II motors are non-processive and cannot produce actin filament sliding or contraction [13,14]. However, they can assemble into processive filaments composed of 10-30 tail-to-tail associated myosins [15]. Sliding driven by myosin filaments can in principle give rise to either a contractile or an extensile force [16,17]. Yet, cellular and reconstituted actin-myosin networks are predominantly contractile. Multiple mechanisms have been identified that bias actin networks towards contraction, including the nonlinear mechanical response of actin filaments to pushing versus pulling forces [18-21] and motor-driven polarity sorting of actin filaments [22-24]. Experiments on cells and reconstituted actin-myosin gels, together with theoretical modelling, furthermore established that the length scale of contraction depends on the network connectivity, which is modulated by the density and length of the actin filaments and the presence of actin-binding proteins that crosslink the network [16,25-30]. Crosslinking is required to help transmit myosin forces over long distances [31], but excess crosslinking can hamper contraction [17,32]. Conversely, myosin motors can actively change the network connectivity by promoting crosslink unbinding and actin filament breakage [20,31,33,34]. Altogether, it is now relatively well understood which parameters govern the length scale of contraction.

By contrast, it remains less clear what governs the dynamics of contraction. Several studies have measured the time scale of macroscopic actin-myosin gel contraction, typically by measuring strain or contractile force as a function of time [20,32,35-38]. By contrast, measurements that probe the microscopic dynamics of contractile networks are relatively scarce. There have been a few microrheology studies that indirectly probed the network dynamics by measuring the fluctuations of embedded probe particles, often using video microscopy [39-42]. These studies have found that motors induce non-thermal fluctuations, which evolve with sample age. However, probe particles are sensitive to the heterogeneous structure of actin-myosin gels because they have a comparable size to the network pores, and they may even locally modify the structure of the gel [43-45]. Furthermore, microrheology studies were made on networks where macroscopic contraction was prevented by choosing appropriate biochemical conditions or by pinning to the boundaries. The microscopic dynamics of synthetic active gels obtained by incorporating self-propelled colloidal particles in a passive network was studied by optical microscopy [46]. However, this study was restricted to systems with relatively weak activity, such that no insight could be gained on the mechanisms leading to the mesoscopic or macroscopic rupture that occurs in highly contractile networks [47,48]. Therefore, it currently remains unclear how contractile active gels evolve over time, and which properties characterize their dynamics.

To address these questions, we quantitatively characterize the dynamics of contractile active gels by measuring the dynamical behavior of reconstituted actin-myosin networks with dynamic light scattering and low magnification video microscopy. We probe the microscopic dynamics of the networks without requiring embedded particles and directly correlate these microscopic dynamics to the macroscopic contraction dynamics of the gels. We reveal several new, previously unanticipated dynamic features. First, we find that despite their complexity, contractile active gels initially exhibit simple aging, reminiscent of that of glassy and jammed passive soft matter. Second, we find evidence for two distinct dynamic precursors of contraction, which mark the end of the aging regime. One precursor manifests itself as apparent rejuvenation, with a slow acceleration of the network restructuring dynamics before contraction. The other precursor is evidenced by sudden discrete rearrangement events characteristic of heterogeneous dynamics. Simultaneous measurements at several scattering vectors reveal that the dynamics associated to these sudden rearrangements are length-scale independent. Both precursors are measurable long before macroscopic gel contraction occurs, in striking analogy to the dynamic precursors of macroscopic failure recently unveiled in passive gels subject to a mechanical load [49]. Previous research [50-52] highlighted the analogy between the microscopic dynamics of biological networks and glassy systems, in the regime where the stresses acting on the network are sufficiently small to prevent macroscopic failure. Our work extends this analogy to active gels undergoing macroscopic failure, thereby providing a general framework for the failure of soft networks.

## 2 Results

### 2.1 Outline

In this article, we investigate active biomimetic gels reconstituted from purified monomeric actin, myosin motors, and fascin crosslink proteins. We initiate contraction by mixing these constituents in cylindrical cuvettes and warming the system up to room temperature. Once the protein solutions are mixed, actin rapidly polymerizes into a network and the myosin motors start contracting the gel to a dense cluster. Movie 1 shows two examples of macroscopic contraction events. In Section 2.2, we first report experiments on dynamic light scattering of gels as a function of sample age, which quantifies the sample's microscopic dynamics. In particular, we investigate non-contractile gels to establish control conditions. In Section 2.3, we investigate contractile gels, and find evidence for sudden decorrelation events indicative of contraction. In Section 2.4, we introduce quantitative video microscopy to characterize the gel's macroscopic contraction dynamics. In Section 2.5, we combine video with light scattering to simultaneously characterize the contracting gel's macroscopic and microscopic dynamics. In Section 2.6, we show that the microscopic dynamics evolve with sample age in three distinct stages, and we find evidence for the first dynamic precursor to contraction. In Section 2.7, we find evidence for a second dynamic precursor involving sudden discrete rearrangements.

### 2.2 Dynamic light scattering of non-contracting gels

In order to probe the microscopic dynamics of the actin gels, we illuminate the gels through the side with a laser beam (in-vacuo wavelength $\lambda = 532$ nm) and detect the scattered light with a charge-coupled device (CCD) placed at an angle $\theta = 45°$ (Fig. **S1**a) [53]. The detection volume is given by the laser cross-section ($1/e^2$ beam diameter of 1 mm) times the depth of the sample (5 mm). We place a lens in the light path in order to spatially resolve the scattered light and make an image of the scattering volume [43,54], imaging the top of the sample. The images appear as patterns of speckles (Fig. **S1**b), with typically ~600 speckles per image. The detection angle of 45° corresponds to a scattering vector $q = \frac{4\pi n}{\lambda} \sin \frac{\theta}{2} = 12.0$ μm$^{-1}$, where $n$ is the refractive index of the solvent. This wavelength corresponds to a characteristic scale length $d = \frac{2\pi}{q} = 0.52$ μm, which is comparable to the average mesh size of the actin networks (ca. 0.3 μm at the actin concentration we use of 12 μM [55]).

Each imaged speckle pattern corresponds to a microscopic conformation of the sample. If the sample is static, the speckle pattern does not change. By contrast, if the sample is dynamic, the intensity of each speckle fluctuates in time, with a timescale $\tau_0$. This time, which we will refer to as the relaxation time $\tau_0$, is a measure of the time scale at which gel strands move with respect to each other over a distance of the order of $d = 2\pi/q$. In order to quantitatively determine $\tau_0$, we perform an image correlation analysis (see Methods and Fig. **S2**). In short, we compute the degree of correlation $c_\tau(t)$ between pairs of images at times $t$ and $t + \tau$, where $\tau$ is the time lag. We fit the correlation functions $c_\tau(t)$ to a phenomenological stretched exponential function, $c_\tau(t) = A \exp\left[-\left(\frac{\tau}{\tau_0}\right)^\beta\right]$, with $\beta$ and $\tau_0$ the time-dependent stretching exponent and relaxation time, respectively. Importantly, the degree of correlation is evaluated in a reference frame where any mesoscopic drift (e.g. due to contraction) cancels out, by using mixed spatio-temporal correlation functions [43]. For each gel, we determine the relaxation time $\tau_0$ as a function of sample age $t$ over a sample-age window of 7–14 h.

We first examine the relaxation time $\tau_0$ for non-contracting samples, which can be obtained in different ways (Fig. **S1**c). The first non-contracting sample is a passive sample, prepared with myosin motors and fascin crosslinks, but without the chemical fuel, adenosine triphosphate (ATP). As shown in Fig. **S1**c, the relaxation time $\tau_0$ for this sample monotonically rises with sample age, reaching an apparent steady state value of ~10000 s (2.7 h) after about 4 hours. This long microscopic relaxation time reflects a nearly static gel structure, which is consistent with the formation of a predominantly elastic polymer network. The time scale of 4 h over which the relaxation time evolves is significantly longer than observed in rheological measurements of polymerizing actin networks [56], which showed that the shear modulus required ~1 h to reach a steady-state value. It is likely that dynamic light scattering is more sensitive to small, residual rearrangements of network conformations over long times compared to rheometry. Indeed, similar slow dynamics have been reported in prior dynamic light scattering studies of passive actin networks prepared in the absence of motors [57].

We next consider three active, but nevertheless non-contracting, gels. We first consider a sample with excess monovalent salt (KCl), which is known to weaken motor activity because the motors form short minifilaments that are only weakly processive [58-60] and only weakly bind actin [61,62]. As shown in Fig. **S1**c, this sample again exhibits a monotonic increase of the relaxation time $\tau_0$ over a period of several hours (5.6 h) before reaching an apparent steady-state. However, now the relaxation time at steady-state is $\tau_0 \approx 2000$ s, 4-fold lower than for the passive (no-ATP) sample. This difference likely arises because the motors in the presence of ATP and high KCl bind weakly, whereas myosin in the absence of ATP binds strongly in a so-called rigor state. We next examine two samples at low-salt conditions (50 mM KCl), where the myosin motors are highly processive, but contraction is prevented by using crosslink (fascin) densities below the connectivity percolation threshold required for macroscopic contraction [31]. Both at zero and at low (0.24 μM) fascin concentration, the active samples exhibit a monotonic rise of the relaxation time over a period of ~ 2.7 hours to a final value $\tau_0 \approx$ 100 s. This relaxation time is again much lower than for the passive (no ATP) network, and it is also much lower than for the active sample prepared with excess KCl, perhaps due to the higher motor processivity. We conclude that motor activity clearly results in enhanced microscopic dynamics, as evidenced by a decrease in the relaxation timescale $\tau_0$ as compared to that of passive samples.

## 2.3  Dynamic light scattering of contracting gels

We now turn to gels prepared with a crosslink density past the percolation threshold, so they are capable of macroscopic contraction. To probe the dynamics at different length scales, we now detect the scattered light with four charge-coupled devices (CCDs) placed at angles $\theta = 22.5°$, $45°$, $90°$, and $120°$ (Fig. **1**a) that correspond to scattering vectors $q = \frac{4\pi n}{\lambda} \sin \frac{\theta}{2} = 6.0$, 12.0, 22.2, and 27.2 μm⁻¹. The speckle patterns (Fig. **1**b) therefore correspond to length scales $d = \frac{2\pi}{q} = 1.0$, 0.52, 0.28, and 0.23 μm, respectively. As shown in Fig. **1**c, the relaxation time $\tau_0$ evolves with sample age differently at different wave vectors: it fluctuates around a constant value of 5-15 s for the two larger scattering vectors (green and yellow curves), whereas it exhibits a non-monotonic behavior with two sudden drops at $t \approx 3100$ s and $t \approx 7500$ s for the two smaller scattering vectors (blue and purple curves). This non-monotonic behavior is in striking contrast to the nearly monotonic aging seen in Fig. **S1** for the non-contractile samples.

In order to investigate the age-dependent dynamics further, we select five windows of sample age and perform power-law fits to the $q$-dependence of the relaxation time (Fig. **S3**), according to the functional form $\tau_0 \sim q^\nu$. The value of the exponent $\nu$ can be used to determine whether the network strands move diffusively ($\nu = -2$), ballistically ($\nu = -1$), or with no dependence on the probed length scale ($\nu = 0$). Diffusive dynamics have been observed in passive gels and have been assigned to rapid thermal fluctuations of gel strands [63]. Ballistic dynamics have been reported in a wide variety of transiently crosslinked passive gels, including both colloidal gels [64] and (bio)polymer gels (see e.g. [57]). They have been attributed to network remodeling stemming from the relaxation of internal stresses, stored in the sample as the result of the formation of a disordered, out-of-equilibrium structure during gelation [64]. Length scale-independent dynamics ($\nu = 0$) have been reported in soft systems with heterogeneous dynamics, such as a coarsening foam [65], and correspond to sudden rearrangement events that fully decorrelate the contribution of photons scattered by the region where the decorrelation event occurs. We find that the scaling exponent $\nu$ for the contractile gel depends on the sample age window over which we measure. During the two drops of $\tau_0$, we find $q^0$ scaling indicative of sudden rearrangement events on scales exceeding 1 μm, the largest length scale probed by our experiment. In the intervening time windows, we instead find $\nu$ in the range $-0.7$ to $-1$, consistent with the $q^1$ scaling characteristic of ballistic motion.

The sudden rearrangement events may arise from contraction of the entire gel. To ascertain whether contraction is occurring, we plot the intensity $I$ of speckles in the image plane normalized by the initial average intensity $I_0$. As shown in Fig. **S4**, the normalized intensity initially fluctuates about a constant value of 1.0–1.4 for different values of $q$. After a sample age of ca. 6000 s, the intensity begins to increase towards values of 1.7–2.4. This rise can either be due to densification as a result of macroscopic contraction, or to local remodeling of the gel such that it accommodates more density inhomogeneities at the length scale probed by the scattering experiment. We note that the increase of the scattering intensity is highest for the lowest $q$. This indicates the formation of large, dense clusters of material, which scatter light more efficiently at low $q$. It is therefore indeed likely that contraction occurred. In order to determine directly whether and how contraction occurs, we turn to real-space video imaging.

## 2.4 Macroscopic dynamics probed by real-space video imaging

Real-space imaging allows for direct visualization of the gel, allowing us to determine whether, how and when contraction occurs. We therefore developed an experimental setup (Fig. **2**a) where one CCD positioned at 45° records Fourier-space scattering (Fig. **2**b), while another CCD at 90° simultaneously records real-space video (Fig. **2**c). Both laser light and white light are used to illuminate the sample. When we observe the samples by macroscopic time-lapse imaging with the 90° camera, we find that contractile gels with identical biochemical composition can exhibit two distinct contractile behaviors. Figure **2**d depicts what we call a "free contraction" event (cf. Movie 1). Contraction begins at the bottom surface and proceeds upward, as shown in the *y-t* projection in Figure **2**e. In this case, the gel is only anchored to its upper interface with air. By contrast, Figure **2**f depicts what we call a "pinned contraction" event. Here, the gel is anchored both to the top and bottom surfaces. As a result, the contraction event is delayed, occurring only 12000 s after gel preparation (Fig. **2**g). In the moments before the contraction event, the gel appears macroscopically undeformed. However, stresses have evidently built up across the gel before contraction sets in, as evidenced by the rapid deformation that occurs in the initial moments of the contraction event. Note that we do not have experimental control of boundary adhesion in our experiments. Identical experimental preparation can yield either free or pinned boundary conditions.

In order to quantify the macroscopic deformation of the gels, we compute the contraction ratio $\lambda^{-1} = \frac{l_0}{l}$ (which is the inverse of the elongation ratio $\lambda$), where $l$ is the instantaneous gel length and $l_0$ the initial length, both obtained from the thresholded *y-t* projections (Fig. **2**e,g). As contraction proceeds, the contraction ratio increases from 1 to values close to 2 (Fig. **2**h), indicating volume changes of up to ≈8-fold, assuming isotropic contraction. The slope of the contraction ratio vs. time plot gives the contraction rate, which is nearly threefold higher for the pinned contraction event (17 x $10^{-5}$ s$^{-1}$, red curve) than for the free contraction event (6 x $10^{-5}$ s$^{-1}$, orange curve, see additional data in Fig. S**5**). In dimensional units, these rates correspond to velocities in the 0.01 to 0.1 µm s$^{-1}$ range, depending on the position in the network.

## 2.5 Microscopic dynamics probed by space-resolved dynamic light scattering

In the previous section, we established with video imaging that macroscopic contraction indeed occurs. We now complement these data to the information gained from concurrent dynamic light scattering measurements.

First, we measure the drift velocity $v_{\text{drift}}$ of speckles in the image plane (Fig. **3**a and Fig **S6**). Drift of speckles corresponds to drift in the sample parallel to the imaged plane, due to the imaging geometry of the setup. Using mixed spatio-temporal intensity correlation functions as detailed in Ref. [43], we are able to measure the drift. Fig **3**a shows the time-course of $v_{\text{drift}}$ for three different samples. For a non-contracting sample that contains motors and ATP but no crosslink protein (blue curve), we find a flat curve with small fluctuations around an average baseline value of around 0.05–0.1 μm / s, which corresponds to spurious drift due to noise measured by the algorithm. By contrast, for the pinned and free contraction samples (orange and red curves, respectively), we find peaked curves. The peaks begin at baseline values in the same range as the non-contracting sample. The peaks then rise to a maximum drift velocity of ~0.4 μm s$^{-1}$ for the free contraction sample and ~0.2 μm s$^{-1}$ for the pinned contraction sample. The curves finally return to baseline values. Not all contractile samples demonstrate clear, uniform drift. The peak values of the drift velocity in the 0.2–0.4 μm s$^{-1}$ range are consistent with the 0.01–0.1 μm s$^{-1}$ range measured from the $y$-$t$ projections of the video images in Fig. **2**.

Second, we measure the average scattering intensity $I$ of speckles in the image plane normalized by the initial average intensity $I_0$. As shown in Fig. **3**b and Fig. **S7**, $I/I_0$ fluctuates about a value of 1 for the non-contracting sample (blue curve), consistent with the observation that this sample does not contract and hence its structure is essentially unchanging. By contrast, $I/I_0$ increases gradually from 1 to 1.5 over approximately 14000 s and then turns up to values of 3-7 for the pinned and free contraction samples (orange and red curves). These values are in reasonable agreement with the ~8-fold macroscopic volume change estimated from video. This agreement suggests that at the probed length scale, the contribution from macroscopic contraction dominates over any microstructural changes. The initial, slow increase in intensity may be due to coarsening of the network. It is known that motors tend to organize actin filaments into dense foci with sizes in the range of 2–20 μm, comprised of a myosin core and actin coat [16,18,24,31].

We were thus able to link both the (Fourier-space) drift velocity and intensity of speckles to the (real-space) volume changes associated with the contraction ratio $\lambda^{-1}$. We therefore from here on define the time of contraction $t_{contraction}$ as the sample age when the normalized scattering intensity exceeds 1.5 (vertical dashed line in Fig. **3**b). Strikingly, $t_{contraction}$ coincides with the peaks in drift velocity in Fig. **3**a, suggesting that our definition indeed faithfully captures the moment when contraction is detected. The scattering intensity trajectories of the pinned (red curve) and free (orange curve) contraction samples are similar, even though their macroscopic strain trajectories observed by video microscopy are markedly different (cf. Fig **2**e). This discrepancy may be explained by the fact that contraction propagates as a wave, from the bottom of the sample where the gel detaches, toward the top where the gel is anchored. Because we measure scattering only at the top of the sample, the local measurement of scattering intensity only increases once the contraction wave reaches the measurement volume. Therefore, our definition of $t_{contraction}$ only captures whether the portion of the sample that is illuminated by the laser is contracting.

## 2.6 Three stages of sample evolution

We now turn to the temporal evolution of the relaxation time of the contractile samples from Figs. **2** and **3**. Fig. **4**a compares the time trajectories for two contractile gels with an identical biochemical composition that exhibit either free or pinned contraction (orange and red lines, respectively). In contrast to the monotonic trajectory of non-contracting gels (blue line; data from Fig. **S1**c.), the relaxation time for the two contractile samples follows a non-monotonic trajectory. The relaxation time first rises to a maximum value of ≈10 s, then falls to a trough of 3–5 s, and finally rises again. This non-monotonic trajectory is reminiscent of the time trajectory for the contractile sample depicted in Fig. **1**, where we also observed sudden drops in $\tau_0$. Furthermore, the relaxation time is an order of magnitude smaller for the contracting gels than for the noncontractile sample. Interestingly, the relaxation time scales we observe for contractile samples are consistent with the time scale of motor-induced active fluctuations observed in microrheology studies, which were in the frequency range of 0.1–10 Hz [39,66].

The non-monotonic time course of the microscopic relaxation time observed for the contracting gels can be broken down into at least three distinct stages, which we label as follows (Fig. **4**a):

<u>I. Aging</u>: Right after sample preparation, $\tau_0$ increases with sample age. As shown in Fig. **4**b, this initial regime for the contractile samples (red and orange curves) is nearly identical to the aging behavior of the non-contracting samples (blue low-fascin and no-fascin curves from Fig. **S1**). We call this first stage, where the microscopic structure becomes increasingly frozen, the *aging* stage. Fig. **4**b displays a log-log plot of the relaxation time $\tau_0$ versus sample age $t$ multiplied by a rescaling factor $\alpha$. We find that the time dependencies exhibit a power law functional form with an average exponent of 1.2±0.3 across all samples. This exponent is consistent with the simple aging reported for passive, glassy and jammed systems [67,68]. Simple aging is defined as a linear relationship between relaxation time and sample age, and as its name indicates, offers the simplest functional form that describes the dynamics of materials with evolving microscopic structure. Interestingly, we find simple aging also for active gels, both for contractile and non-contracting samples.

<u>II. Rejuvenation</u>: For contractile samples, $\tau_0$ reaches a peak value in the range of 10–20s during the aging stage and next decreases again towards a minimum with values of 1–5s. We denote this second stage as the *rejuvenation* stage because it is characterized by increased structural rearrangements that are presumably caused by motor activity. Indeed, we recall that samples where motor activity is inhibited or absent do not exhibit rejuvenation. Interestingly, the sample age where $\tau_0$ reaches a minimum coincides with the contraction time $t_{contraction}$ where the scattering intensity starts to increase (cf. Fig **3**d). Surprisingly, the rejuvenation stage can initiate well before contraction is detected, in one sample even up to 10000s (~2.8h) prior (Fig. **S8**, light orange curve). The smooth decrease of $\tau_0$ during the rejuvenation stage therefore amounts to a dynamic precursor of macroscopic contraction.

<u>III. Contraction</u>: Once the gels start to macroscopically contract, the microscopic structure appears to become more static again, as indicated by an increase of $\tau_0$ towards values of 10–30 s. Possible explanations are that the gel becomes denser from contraction and/or that the contraction event releases internal stresses. Interestingly, this slowing down of the dynamics upon densification is reminiscent of the suppression of cytoskeleton remodeling in living cells subjected to osmotic compression [51].

## 2.7 Pinned contractile gels exhibit discrete stress-driven rearrangements

The video data showed that pinned contraction samples do not contract immediately. This is evident in Fig $\mathbf{2}$g, which reveals a significant lag time before contraction, during which stresses apparently build up. This raises the question whether the microscopic dynamics of a pinned gel undergoing stress buildup differ from a freely contracting gel. To answer this question, we consider the time dependence of the degree of correlation $c_\tau(t)$ between speckle images for lag times $\tau$ close to the typical relaxation time $\tau_0$. In particular, we look at the fluctuations of the dynamics as quantified by $c_{6s} - <c_{6s}>_{100s}$, where $c_{6s}$ is the degree of correlation for a lag time $\tau = 6s$, and $<\bullet>_{100s}$ denotes a temporal average over a 100s window. We chose the timescale of 6s to agree roughly with the relaxation time of the contractile samples, and the timescale of 100s to filter out the long-time evolution of the degree of correlation.

Figure $\mathbf{5}$a shows $c_{6s} - <c_{6s}>_{100s}$ for a non-contracting active gel (prepared with motors but without crosslinkers), which mainly exhibits fluctuations about zero. We attribute these fluctuations to statistical noise stemming from the finite number of speckles in the CCD images [69]. Figure $\mathbf{5}$b shows $c_{6s} - <c_{6s}>_{100s}$ for a freely contracting gel. Apart from fluctuations about zero, this sample exhibits dips near 4000s indicative of discrete decorrelation events (Fig. $\mathbf{5}$b). These dips occur even more prominently in an active gel undergoing pinned contraction (Fig. $\mathbf{5}$c). They occur in stages I and II, before the onset of macroscopic contraction, but not in stage III, during contraction. These decorrelation events last only 1–10 s, which is probably why they do not significantly affect $\tau_0$.

The decorrelation events can also be quantified by considering the histogram of the fluctuation values (Fig. $\mathbf{5}$d). The histogram for the non-contracting sample (blue curve) has a Gaussian shape and low variance ($\sigma^2 = 0.0003$), characteristic of temporally homogeneous dynamics and stemming from the statistical noise discussed above [69]. By contrast, the contractile samples (orange and red curves) exhibit non-Gaussian distributions indicating temporally heterogeneous dynamics [69-72], with long tails towards negative values characteristic of decorrelation events and a high variance ($\sigma^2 = 0.0015$–$0.003$). The variance of the distribution is indicative of the dynamic susceptibility $\chi_4$ [69], which is widely used as an indicator of dynamic heterogeneity in glassy systems [73]. The variance has the largest values during stage II (rejuvenation) and lowest values in stage III (contraction) for different contractile samples (Figs. $\mathbf{S9}$ and $\mathbf{S10}$). These results show stress buildup before contraction is determined by discrete stress-driven rearrangements, which may include depinning of the gel from the boundaries and internal network rupture.

# 3 Discussion

To reveal the microscopic dynamics that govern the contractile activity of active cytoskeletal gels, we have performed simultaneous real-space video and Fourier-space light-scattering measurement on reconstituted actin networks driven by myosin motors. We find that the active gels show rich microscopic dynamics that evolve with sample age in three distinct stages.

We showed that the samples initially age (stage I), as evidenced by a linear increase of the microscopic relaxation time $\tau_0$ with sample age (cf. Fig. **4**b). This linear dependence demonstrates that active gels exhibit simple aging. This result is not immediately expected because there are several competing processes that influence stress build-up during the aging stage, including actin polymerization, actin filament entanglement, crosslink binding and unbinding, and crosslink-mediated bundle formation kinetics [57,74,75]. In addition, active samples (both contractile and non-contractile) are expected to exhibit motor-induced sliding of actin filaments [76]. Understanding how all these processes contribute to the simple aging behavior will require combining different microscopic theories and computational frameworks that currently model these processes separately [17,74,77-79].

Whereas the aging stage (stage I) is marked by multiple competing processes, the rejuvenation stage (stage II) is dominated by myosin motor activity. We find evidence for two motor-induced relaxation mechanisms that are unique to the contractile samples and reflect dynamic precursors to macroscopic contraction. The first mechanism is a continuous relaxation mechanism, characterized by a gradual decrease of $\tau_0$ over prolonged periods of up to 10000 s before contraction (cf. Fig. **4**a and **S8**, left sides of plots). This behavior is reminiscent of failure precursors observed in passive colloidal gels submitted to a constant shear stress, which also manifested as enhanced microscopic dynamics before macroscopic failure [49]. By analogy, we interpret the enhanced dynamics of actin-myosin networks as a dynamic precursor to macroscopic contraction. By contrast, the second relaxation mechanism that occurs during the rejuvenation stage and is superposed on the enhanced dynamics is characterized by sudden decorrelation events. These events are too short to significantly affect $\tau_0$ but they show up as strong negative peaks in the degree of correlation (cf. Fig. **5**b–d) and as non-Gaussian histograms of the degree of correlation (cf. Fig. **5**d and S**9**a). These heterogeneous dynamics were particularly prominent in samples exhibiting pinned contraction, where stress builds up for a long time before contraction occurs (cf. Fig. **2**g). We therefore propose that these discrete remodeling events involve local depinning of the actin network from anchoring surfaces as well as local nucleation and propagation of ruptures within the material identified in previous experimental and computational studies of contractile gels [31,48,80,81]. These rupture events likely gradually weaken the network before it ultimately collapses. Prior studies have investigated dynamic precursors to catastrophic failure events across a broad range of systems, including earthquakes [82], avalanches [83], and fracture of gels [49,84]. The dynamic precursors to motor-induced contraction which we identify here demonstrate that similar events occur in active systems that evolve by internal driving.

To probe the length scale dependence of the dynamics, we also measured the microscopic dynamics of a typical contraction event simultaneously at different $q$-vectors (cf. Fig. **1**). We found different $q$-dependencies of the relaxation time $\tau_0$ depending on sample age. We measured a scaling exponent of −1 during the aging stage (stage I, window 2000–2500 s) and the contraction stage (stage III, window 8000–9000 s). This exponent is consistent with prior studies of passive systems, including actin networks cross-linked with fascin [57] and fractal gels of polystyrene colloids [71,85]. The exponent of −1 in these systems was attributed to slow relaxation of internal stress and glassy dynamics. During the rejuvenation stage (stage II), we find a tendency to slightly lower (in absolute value) exponents, $\nu = -0.7$. We interpret this observation as follows. Stage I is likely determined by actin filament and crosslink dynamics, so the dynamics of stage I resemble that of a passive gel. In stage II, effects from myosin motors arise, both in the continuous and discrete relaxation mechanisms mentioned above. These active effects are likely to decrease the magnitude of the exponent due to a crossover effect between the $\nu = -1$ value of regime I and the $\nu = 0$ value observed during major decorrelation events (see below). Once the sample has begun to macroscopically contract in stage III, the dynamics are dominated by the contractile strain rate. In this regime, the macroscopic dynamics follows the macroscopic deformation, as seen in passive gels [49,86], for which ballistic dynamics ($\nu = -1$) was reported.

As it ages, the contractile sample undergoes sudden major decorrelation events (at $t = 3100$ s and $t = 7400$ s in Fig. **1**). During these events, the scaling exponent $\nu$ is close to zero, implying dynamics independent of the length scale that is probed. This striking behavior can be rationalized as the result of discrete, localized rearrangement events that reconfigure the network structure by displacing gel strands over distances larger than the largest length scale probed by our light scattering experiments, of the order of 1 μm. Indeed, light scattered by a rearranged region is completely decorrelated at all probed $q$ vectors. Therefore, the decay rate of the correlation functions is identical for all scattering vectors, being only dictated by the rearrangement rate per unit volume. Length-scale-independent dynamics are quite rare. They have been reported for a shaving cream foam, where they stem from sudden bubble rearrangements resulting from internal stress that builds up during foam coarsening [65]. They have also been observed in passive colloidal gels, where they have been attributed to rearrangement events triggered by internal stress resulting from the rapid, disordered gelation process [71]. They have furthermore been found in DNA tetravalent networks, where they have been attributed to fluctuations in local elasticity and connectivity [87].

In passive gels, length-scale-independent dynamics are difficult to observe, because individual rearrangement events involve displacements comparable to thermally activated fluctuations of the network at fixed connectivity [71]. By contrast, these events are clearly seen in the active gels studied here, showing that the stress build-up generated by the motors has a major impact on the network dynamics, entailing bond breaking and network remodeling on length scales much larger than those corresponding to the thermal fluctuations at fixed network connectivity. More generally, our investigation unveils that the microscopic dynamics of contractile active gels involve precursor plastic rearrangements that progressively weaken the network, eventually triggering macroscopic contraction.

# 4 Conclusion

We investigated the macroscopic and microscopic dynamics of contractile active gels driven by molecular motor activity. We measured the macroscopic state of the gels with time-lapse video imaging, and simultaneously measured the corresponding microscopic dynamics using space- and time-resolved dynamic light scattering. We uncovered three dynamical properties. First, we found that active gels initially exhibit simple aging (stage I). Second, we found that aging is followed by a self-rejuvenation stage (stage II), characterized by a gradual decrease of the relaxation time $\tau_0$. Third, we found an increased occurrence of decorrelation events and hence heterogeneous dynamics. These observations point to two separate stress-relaxation mechanisms: a continuous one and a discrete one. The latter is associated to an anomalous scaling of the relaxation time $\tau_0$ with the scattering vector $q$, which we attribute to the occurrence of sudden displacements spanning length scales greater than 1 μm, the largest length scale probed by our setup. The dynamics during the self-rejuvenation stage amount to dynamic precursors to motor-driven contraction of active gels. Our findings provide a more detailed understanding of the unusual dynamical properties of active gels driven by molecular motors, and establish an intriguing analogy between the self-induced failure of active, contractile networks and that of passive gels subject to an external load.

# 5 Acknowledgements

We acknowledge T. Divoux for insightful discussions. We gratefully acknowledge financial support from the CNES (LC), a Vidi grant from the Netherlands Organization for Scientific Research (GHK), and the US Army Research Office under grant number W911NF-14-1-0396 (JA). We thank M. Kuit-Vinkenoog for help with protein purifications and S. Hansen and R.D. Mullins (UCSF, San Francisco, USA) for the fascin plasmid.

# 6 Methods

## 6.1 Protein Purification

Monomeric (G-) actin was purchased from Cytoskeleton Inc. in the form of lyophilized powder and resuspended in water, yielding buffer conditions 5 mM Tris-HCl pH 8.0, 0.2 mM $CaCl_2$, 0.2 mM ATP, 5% (w/v) sucrose and 1% (w/v) dextran. Actin solutions were stored on ice (0°C). Myosin II was purified from rabbit psoas skeletal muscle according to a published procedure [18] and stored at −20°C in a high-salt storage buffer with glycerol (25 mM monopotassium phosphate pH 6.5, 600 mM potassium chloride, 10 mM ethylenediaminetetraacetic acid, 1 mM dithiothreitol, 50% w/w glycerol). Creatine phosphate disodium and creatine kinase were purchased from Roche Diagnostics (Indianapolis, IN, USA), all other chemicals from Sigma Aldrich (St. Louis, MO, USA). Magnesium adenosine triphosphate was prepared as a 100 mM stock solution in 10 mM imidazole pH 7.4 using equimolar amounts of disodium adenosine triphosphate and magnesium chloride. Recombinant mouse fascin was prepared from T7 pGEX E. coli [88] using a plasmid provided by Scott Hansen and R. Dyche Mullins (UC, San Francisco). We snap-freeze aliquots of fascin protein and store at –80 °C in 20 mM imidazole pH 7.4, 150 mM potassium chloride, 1 mM dithiothreitol, and 10% v/v glycerol.

## 6.2   Dynamic Light Scattering Experiments

**Sample Preparation.** Active gels were prepared with a fixed actin concentration, [actin] = 12 μM, and with molar ratios of [myosin] / [actin] = $R_M$ = 0, 0.002, 0.005, 0.01 and [fascin] / [actin] = $R_F$ = 0, 0.02, 0.05. Networks were formed in a polymerization buffer composed of 20 mM imidazole (pH 7.4), 50 mM KCl, 2 mM MgCl$_2$, 0.1 mM ATP, 10 mM creatine phosphate disodium, 0.1 mg/mL creatine kinase, and 1 mM trolox. Note that creatine phosphate and creatine kinase are used to regenerate ATP, while trolox is present to suppress photobleaching. All buffer solutions were filtered through 0.22-μm filters to remove dust, which may interfere with light scattering measurements. Networks were prepared by mixing a solution containing buffers, salts, myosin, and fascin; polymerization of the network was initiated by mixing this solution with actin monomers. This mixture was subsequently loaded into a cylindrical glass NMR tube (Spectrometrie Spin et Techniques, Champs-sur-Marne, France) with a 5-mm outer diameter and a wall thickness of 0.4 mm. We define the moment when the solutions are mixed to be sample age $t$ = 0.

**Experimental setup**. We used a setup with up to four charge-coupled devices (CCDs) to measure space- and time-resolved scattering of an incident 532 nm laser collimated to a beam width of 1 mm [69]. The sample is mounted in a temperature-controlled copper housing with four holes drilled at 22.5°, 45°, 90°, and 120°. The four CCDs were mounted to capture speckle dynamics simultaneously at all four angles. Each CCD delivers space- and time-resolved information, and combining all four CCDs allows us to investigate length-scale dependent dynamics. The intensity of each pixel of the CCD corresponds to a scattering volume of 3μm x 3μm x 1mm. The value of 3 μm is determined by the size of the image projected on the CCD and was chosen to coincide with the average speckle size. The value of 1 mm is determined by the width of the beam. Simultaneous video recording and light scattering was performed using another setup equipped with one CCD mounted at an angle of 45° for capturing speckle dynamics and a second CCD at 90° for capturing real-space video.

CCD data acquisition was performed using a variable delay between images, as described previously [89].  In brief, we acquired pairs of images: the delay between two pairs of images varied from 5 ms to 1 s using ¼-order-of-magnitude steps, scaled logarithmically (e.g. 100 ms, 178 ms, 316 ms, 562 ms, 1 s); the delay between subsequent pairs was fixed at 1s. We cycled through these logarithmically separated pairs of images continuously throughout the entire experiment [89]. The acquisition of all CCDs was triggered simultaneously. The image exposure time was 1 ms. The CCD data were processed to correct for dark background and uneven illumination as described previously [69].

The CCD images have a speckled appearance, due to the interference between photons scattered by the sample (see Fig. 2a). With time, the speckle pattern drifts, reflecting the local drift motion of the sample, e.g. due to contraction. Additionally, the intensity of a given speckle fluctuates as a result of the relative motion of the gel strands contributing to that speckle. We quantify the drift motion and the relative dynamics by a mixed spatio-temporal correlation function [43]:

$$c_\tau(t, \Delta x, \Delta y) = \frac{\langle I(x,y,t) I(x + \Delta x, y + \Delta y, t + \tau) \rangle}{\langle I(x,y,t) \rangle \langle I(x + \Delta x, y + \Delta y, t + \tau) \rangle} - 1$$

Here, $I(x, y, t)$ is the CCD intensity of a pixel with spatial coordinates $(x,y)$ at time $t$ and <...> denotes an average over a rectangular region of interest (ROI) centered around $(x,y)$. The ROI size was typically 235 x 107 pixels.

The spatio-temporal correlation function exhibits a peak, whose position yields the drift between times $t$ and $t+\tau$ while the height quantifies the microscopic dynamics in a reference frame co-moving with the local sample drift [43]. To avoid overburdening the notation, in the main text we denote $c_\tau(t, \Delta x^*, \Delta y^*)$ simply by $c_\tau(t)$. For a sample with no drift and stationary dynamics, the $t$-averaged $c_\tau$ reduces to the usual intensity correlation function $g_2(\tau) - 1$ measured in conventional dynamic light scattering [90].

**Correlation analysis**: We perform correlation analysis on pairs of speckle images acquired by the CCDs. First, we consider a pair of images, taken at times $t$ and $t + \tau$. We denote $t$ as the *sample age*. We define $t = 0$ as the time where we initiate actin polymerization. We denote τ as the *lag time*, and restrict it to positive values. Next, we compute the *degree of correlation $c_\tau(t)$*, which takes values between 0 and a setup-dependent constant $A \lesssim 1$. A value of 0 corresponds to two completely unrelated images, whereas a value of $A$ of 1 occurs when the two images are identical. See Fig. **S2**a–f for examples. Next, we compute $c_\tau(t)$ for the first image at $t = 0$ and an array of images corresponding to different values of τ. Rather than analyzing images for all possible values of τ, we restrict ourselves to ¼-order-of-magnitude steps, scaled logarithmically (e.g. 1 s, 2 s, 4 s, 6 s, 10 s, …). We plot $c_\tau(t)$ as a function of $\tau$ and fit the data to a stretched exponential:

$$c_\tau(t) = A \exp\left(-\left(\frac{\tau}{\tau_0}\right)^\beta\right)$$

The fit produces three constants: the amplitude $A$, the stretching exponent $\beta$, and the relaxation timescale $\tau_0$. Repeating this analysis for different values of the sample age $t$ yields trajectories $\tau_0(t)$ which describe the evolution of the microscopic sample dynamics. Rather than performing the analysis for all possible values of $t$, we group values of $c_\tau(t)$ into intervals of a certain duration $\Delta t$ and average within these intervals.

# 7  Figures

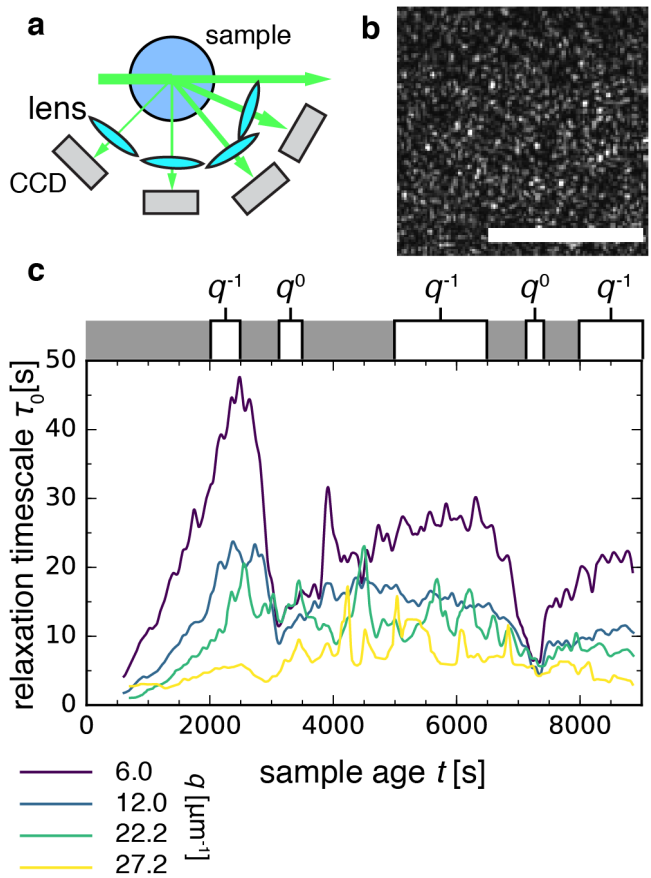

**Figure 1**. Length scale dependence of the microscopic dynamics of a typical contractile sample reveals $q$-dependent and $q$-independent episodes. The sample contains [actin] = 12 μM, [fascin] = 0.6 μM, [myosin] = 0.12 μM, and [ATP] = 0.1 mM. (a) Schematic of light scattering setup, top view. Scattered light is acquired at four different angles corresponding to scattering vectors $q$ = 6.0, 12.0, 22.2, 27.2 μm⁻¹. (b) Typical speckle pattern recorded by the CCD at 45°. Full images contain ca. 500–700 speckles. Scale bar 200 μm. (c) Time course of the relaxation time $\tau_0$ for the four scattering vectors. Gray bar and white boxes denote time intervals over which a power-law fit was performed on $\tau_0$ vs $q$ data. See Fig. S**3** for precise exponents and plots of power-law fits. Bin size $\Delta t$ = 10 s.

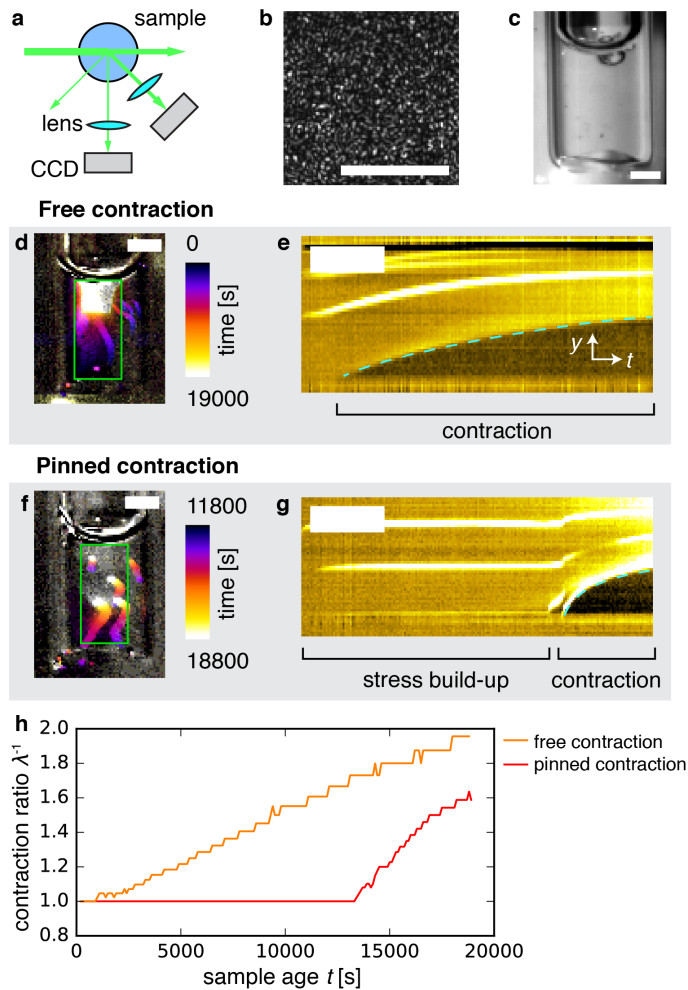

**Figure 2.** Macroscopic dynamics of contractile active gels, as measured by real-space imaging. (Also see supplemental Movie 1.) (a) Schematic of the experimental setup, top view. The sample is contained in a cylindrical tube. A laser beam (in-vacuo $\lambda$ = 532 nm) passes through the gel. Fourier-space scattered light is recorded at an angle $\theta$ = 45°, while real-space video is recorded at $\theta$ = 90°. (b) Typical speckle pattern recorded by the CCD. Full images contain ca. 500–700 speckles. Scale bar 200 μm. (c) Image of the cuvette containing the active gel at $t$ = 0. Scale bar 2 mm. (d) Color-time overlay of an active gel that contracts freely. Color corresponds to time (calibration bar, right). Scale bar 2 mm. (e) $y$-$t$ representation of green box from panel a. The vertical direction denotes the spatial $y$-direction (scale bar 1 mm), while the horizontal direction denotes time (scale bar 4000 s). The $x$-direction of the green box is mean-projected. The dashed cyan line is a guide to the eye, which tracks the bottom edge of the gel. (f) Color-time overlay of an active gel that is pinned at the top and bottom surfaces, as in panel d. (g) $y$-$t$ representation of the green box from panel f, as in panel e. (h) Contraction ratio $\lambda^{-1}$ as a function of sample age $t$, based on the $y$-$t$ projections in panels e and g for free (top, orange) and pinned (bottom, red) contraction events. The two samples have an identical biochemical composition ([fascin] = 0.6 μM, [myosin] = 0.06 μM, [actin] = 12 μM, [ATP] = 0.1 mM). The two curves quantify the time evolution of the dashed cyan lines in panels e and g.

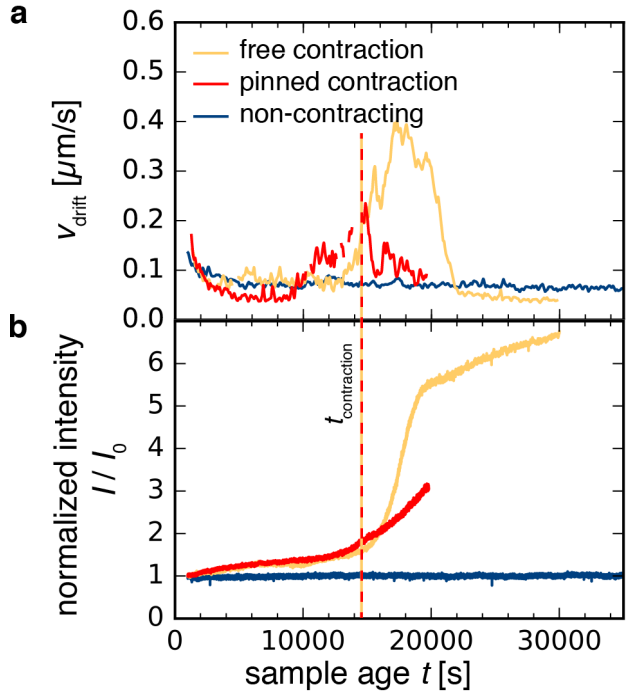

**Figure 3.** Microscopic dynamics of contractile and non-contracting active gels, as measured by space-resolved dynamic light scattering. (a) Drift velocity $v_{drift}$ of speckles in the imaging plane of the CCD as a function of sample age $t$, for two active contractile gels ([fascin] = 0.6 μM, [myosin] = 0.06 μM, [actin] = 12 μM, [ATP] = 0.1 mM) that exhibit free contraction (orange line) or pinned contraction (red line), and for one active, but non-contracting sample (blue line; [fascin] = 0 μM, [myosin] = 0.06 μM, [actin] = 12 μM, [ATP] = 0.1 mM). (b) Scattering intensity $I$ normalized by the initial intensity $I_0$ as a function of sample age $t$. The vertical dashed line indicates the sample age where contraction sets in, as defined by the increase in normalized scattering intensity past the value of 1.5. Bin size $\Delta t$ = 10 s for the free and pinned contraction samples, 50 s for the non-contracting sample.

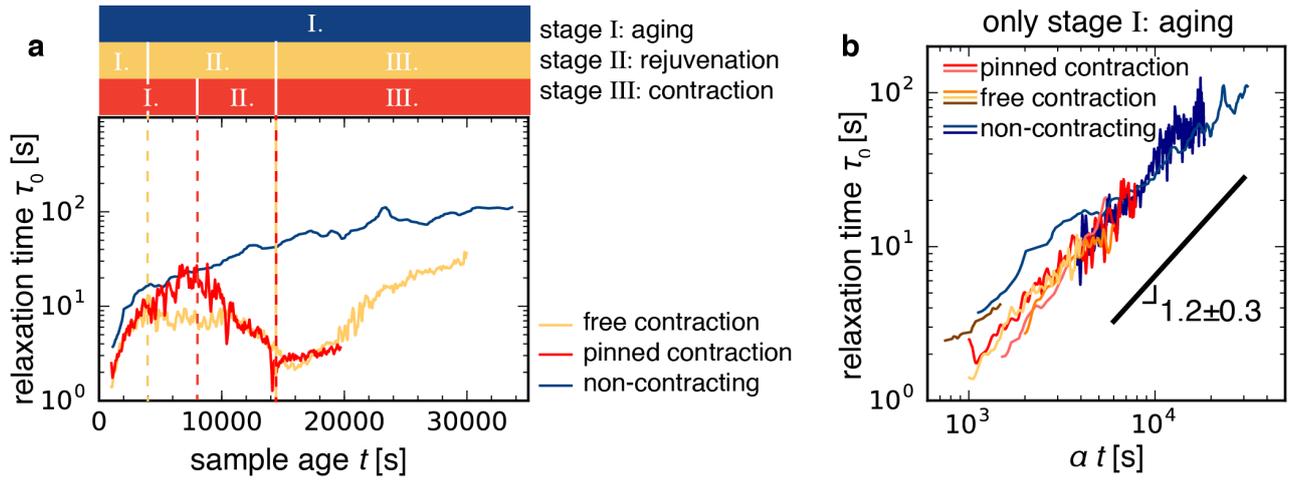

**Figure 4.** The time evolution of the relaxation time $\tau_0$ reveals non-monotonic microscopic dynamics for active contractile gels. (a) Time trajectories for the three samples shown in Fig. **3**. Colored bars denote the three stages of contractile gel evolution as defined in the main text. Stage I is aging, where $\tau_0$ increases with sample age. Stage II is rejuvenation, where $\tau_0$ decreases with sample age. Stage III is the contraction stage, where $\tau_0$ starts to increase right after contraction begins. The time dependencies are comparable for the networks exhibiting free or pinned contraction, suggesting that boundary adhesion does not significantly affect $\tau_0$. The non-contracting gel (blue line) does not exhibit rejuvenation nor contraction. (b) Log-log plots of relaxation time $\tau_0$ as a function of sample age $t$ during stage I collapse onto a single master curve upon rescaling with a factor $\alpha$, which varies in the range 0.5–2.0. The black line denotes a power-law with the best-fit exponent 1.2 ± 0.3 obtained from averaging over fits to individual curves. Red-orange lines denote contractile samples; blue lines denote passive and active, but non-contracting, samples.

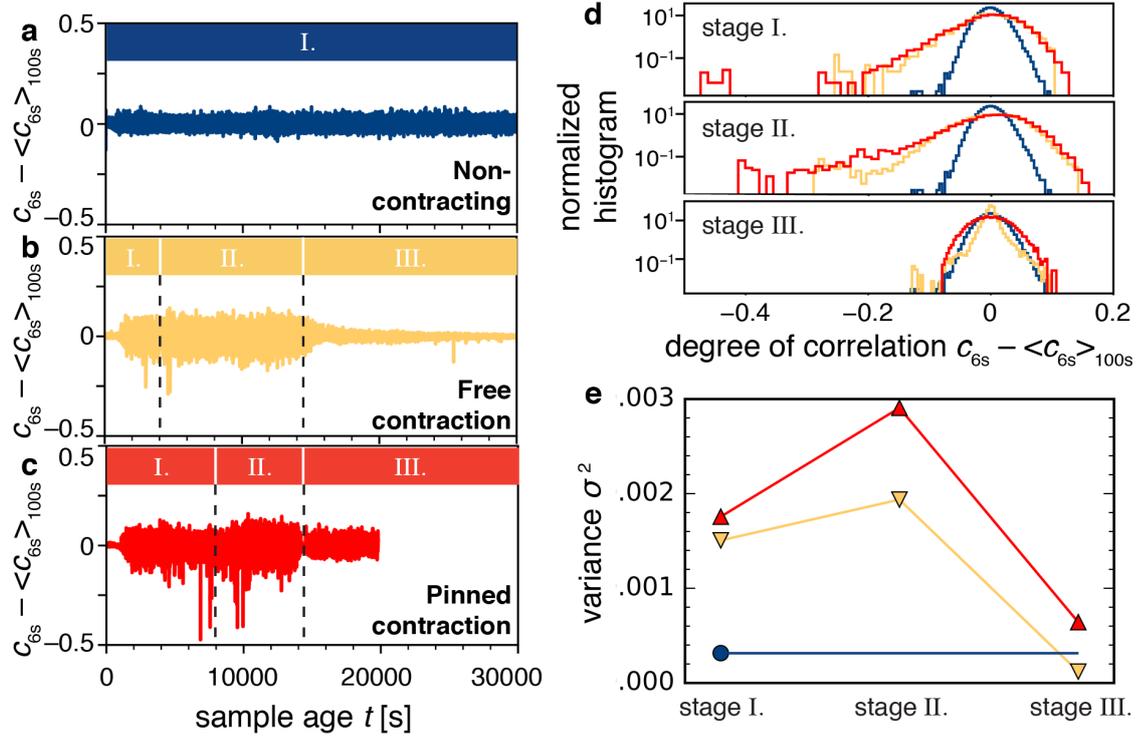

**Figure 5.** Decorrelation events result in discrete stress relaxation, which is enhanced for gels exhibiting pinned contraction over gels exhibiting free contraction. (a–c) Fluctuations of the degree of correlation, quantified by $c_{6s}$ minus its Gaussian-filtered baseline $<c_{6s}>_{100s}$, as a function of sample age for (a) the non-contracting gel, (b) the freely contracting gel, and (c) the pinned contraction gel from Fig. **3**. Colored bars and black dashed lines denote the aging stage (I), the rejuvenation stage (II), and the contraction stage (III). (d) Histograms of $c_{6s}-<c_{6s}>_{100s}$ from panels a–c broken down by stage. (e) Variance $\sigma^2$ of the distributions of panel d for different stages and different samples: non-contracting (blue circle), free contraction (orange downward triangles), and pinned contraction (red upward triangles).

# 8 Bibliography

[1]     D. Needleman and Z. Dogic, Nature Reviews Materials 2017 **2**, 17048 (2017).

[2]     A. Diz-Muñoz, O. D. Weiner, and D. A. Fletcher, Nature Phys **14**, 648 (2018).

[3]     G. H. Koenderink and E. K. Paluch, Curr Opin Cell Biol **50**, 79 (2018).

[4]     M. Murrell, P. W. Oakes, M. Lenz, and M. L. Gardel, Nat Rev Mol Cell Bio **16**, 486 (2015).

[5]     J. Howard, *Mechanics of Motor Proteins &Amp; the Cytoskeleton* (Sinauer Associates, 2001).

[6]     P. Roca-Cusachs, V. Conte, and X. Trepat, Nature Cell Biol **19**, 742 (2017).

[7]     M. Gautel, Curr Opin Cell Biol **23**, 39 (2011).

[8]     F. Spira, S. Cuylen-Haering, S. Mehta, M. Samwer, A. Reversat, A. Verma, R. Oldenbourg, M. Sixt, and D. W. Gerlich, eLife **6**, 983 (2017).

[9]     M. Vicente-Manzanares, J. Zareno, L. Whitmore, C. K. Choi, and A. F. Horwitz, J Cell Biol **176**, 573 (2007).

[10]    M. Mayer, M. Depken, J. S. Bois, F. Jülicher, and S. W. Grill, Nature **467**, 617 (2010).

[11]    M. Rauzi, P. Verant, T. Lecuit, and P.-F. Lenne, Nature Cell Biol **10**, 1401 (2008).

[12]    R. Fernandez-Gonzalez and J. A. Zallen, Mol Biol Cell **24**, 3227 (2013).

[13]    F. Wang, M. Kovács, A. Hu, J. Limouze, E. V. Harvey, and J. R. Sellers, J Biol Chem **278**, 27439 (2003).

[14]    M. Kovács, F. Wang, A. Hu, Y. Zhang, and J. R. Sellers, J Biol Chem **278**, 38132 (2003).

[15]    X. Liu, N. Billington, S. Shu, S.-H. Yu, G. Piszczek, J. R. Sellers, and E. D. Korn, Proc Natl Acad Sci USA **114**, E6516 (2017).

[16]    S. Stam, S. L. Freedman, S. Banerjee, K. L. Weirich, A. R. Dinner, and M. L. Gardel, Proc Natl Acad Sci USA **114**, E10037 (2017).

[17]    J. M. Belmonte, M. Leptin, and F. Nedelec, Molecular Systems Biology **13**, 941 (2017).

[18]    M. Soares e Silva, M. Depken, B. Stuhrmann, M. Korsten, F. C. Mackintosh, and G. H. Koenderink, Proc Natl Acad Sci USA **108**, 9408 (2011).

[19]    M. Lenz, T. Thoresen, M. Gardel, and A. Dinner, Phys Rev Lett **108**, 238107 (2012).

[20]    M. P. Murrell and M. L. Gardel, Proc Natl Acad Sci USA **109**, 20820 (2012).

[21]    P. Ronceray, C. P. Broedersz, and M. Lenz, Proc Natl Acad Sci USA **113**, 2827 (2016).

[22]    K. Kruse and F. Julicher, Phys Rev Lett **85**, 1778 (2000).

[23]    D. B. Oelz, B. Y. Rubinstein, and A. Mogilner, Biophys J **109**, 1818 (2015).

[24]    V. Wollrab, J. M. Belmonte, L. Baldauf, M. Leptin, F. Nedelec, and G. H. Koenderink, J Cell Sci **132**, jcs219717 (2019).

[25]    L. W. Janson, J. Kolega, and D. L. Taylor, J Cell Biol **114**, 1005 (1991).

[26]    S. Köhler, V. Schaller, and A. R. Bausch, Nature Mater **10**, 462 (2011).

[27]    Y. Ideses, A. Sonn-Segev, Y. Roichman, and A. Bernheim-Groswasser, Soft Matter **9**, 7127 (2013).

[28]    H. Ennomani, G. Letort, C. Guérin, J.-L. Martiel, W. Cao, F. Nedelec, E. M. De La Cruz, M. Théry, and L. Blanchoin, Curr Biol **26**, 616 (2016).

[29]    P. Chugh, A. G. Clark, M. B. Smith, D. A. D. Cassani, K. Dierkes, A. Ragab, P. P. Roux, G. Charras, G. Salbreux, and E. K. Paluch, Nature Cell Biol **19**, 689 (2017).

[30]    W. Y. Ding, H. T. Ong, Y. Hara, J. Wongsantichon, Y. Toyama, R. C. Robinson, F. Nedelec, and R. Zaidel-Bar, J Cell Biol **216**, 1371 (2017).

[31]    J. Alvarado, M. Sheinman, A. Sharma, F. C. MacKintosh, and G. Koenderink, Nature Phys **9**, 591 (2013).

[32]    P. M. Bendix, G. H. Koenderink, D. Cuvelier, Z. Dogic, B. N. Koeleman, W. M. Brieher, C. M. Field, L. Mahadevan, and D. A. Weitz, Biophys J **94**, 3126 (2008).

[33] L. Haviv, D. Gillo, F. Backouche, and A. Bernheim-Groswasser, J Mol Biol **375**, 325 (2008).

[34] A. C. Reymann, R. Boujemaa-Paterski, J. L. Martiel, C. Guerin, W. Cao, H. F. Chin, E. M. De La Cruz, M. Thery, and L. Blanchoin, Science **336**, 1310 (2012).

[35] H. Wang, A. A. Svoronos, T. Boudou, M. S. Sakar, J. Y. Schell, J. R. Morgan, C. S. Chen, and V. B. Shenoy, Proc Natl Acad Sci USA **110**, 20923 (2013).

[36] P. J. Foster, S. Fürthauer, M. J. Shelley, and D. J. Needleman, eLife 2015;4:e10837 (2015).

[37] I. Linsmeier, S. Banerjee, P. W. Oakes, W. Jung, T. Kim, and M. P. Murrell, Nature Commun **7**, 12615 (2016).

[38] M. Schuppler, F. C. Keber, M. Kröger, and A. R. Bausch, Nature Commun **7**, 13120 (2016).

[39] D. Mizuno, C. Tardin, C. F. Schmidt, and F. C. MacKintosh, Science **315**, 370 (2007).

[40] D. Mizuno, D. A. Head, F. C. MacKintosh, and C. F. Schmidt, Macromolecules **41**, 7194 (2008).

[41] B. Stuhrmann, M. Soares e Silva, M. Depken, F. C. MacKintosh, and G. H. Koenderink, Phys Rev E **86**, 020901(R) (2012).

[42] A. Sonn-Segev, A. Bernheim-Groswasser, and Y. Roichman, J Phys Condens Matter **29**, 163002 (2017).

[43] L. Cipelletti, G. Brambilla, S. Maccarone, and S. Caroff, Opt Express **21**, 22353 (2013).

[44] J. He and J. Tang, Phys Rev E **83**, 041902 (2011).

[45] B. S. Chae and E. M. Furst, Langmuir **21**, 3084 (2005).

[46] M. E. Szakasits, W. Zhang, and M. J. Solomon, Phys Rev Lett **119**, 058001 (2017).

[47] M. Sheinman, A. Sharma, J. Alvarado, G. H. Koenderink, and F. C. MacKintosh, Phys Rev Lett **114**, 098104 (2015).

[48] C. F. Lee and G. Pruessner, Phys Rev E **93**, 052414 (2016).

[49] S. Aime, L. Ramos, and L. Cipelletti, Proc Natl Acad Sci USA **115**, 3587 (2018).

[50] P. Bursac, G. Lenormand, Ben Fabry, M. Oliver, D. A. Weitz, V. Viasnoff, J. P. Butler, and J. J. Fredberg, Nature Mater **4**, 557 (2005).

[51] E. H. Zhou, X. Trepat, C. Y. Park, G. Lenormand, M. N. Oliver, S. M. Mijailovich, C. Hardin, D. A. Weitz, J. P. Butler, and J. J. Fredberg, Proc Natl Acad Sci USA **106**, 10632 (2009).

[52] S. Wang and P. G. Wolynes, J Chem Phys **138**, 12A521 (2013).

[53] D. El Masri, M. Pierno, L. Berthier, and L. Cipelletti, J Phys Condens Matter **17**, S3543 (2005).

[54] A. Duri, D. Sessoms, V. Trappe, and L. Cipelletti, Phys Rev Lett **102**, 085702 (2009).

[55] C. F. Schmidt, M. Baermann, G. Isenberg, and E. Sackmann, Macromolecules **22**, 3638 (1989).

[56] Y. Tseng, K. M. An, and D. Wirtz, J Biol Chem **277**, 18143 (2002).

[57] O. Lieleg, J. Kayser, G. Brambilla, L. Cipelletti, and A. R. Bausch, Nature Mater **10**, 236 (2011).

[58] H. Noda and S. Ebashi, Biochim Biophys Acta **41**, 386 (1960).

[59] E. Reisler, C. Smith, and G. Seegan, J Mol Biol **143**, 129 (1980).

[60] B. Kaminer and A. L. Bell, J Mol Biol **20**, 391 (1966).

[61] B. Brenner, M. Schoenberg, J. M. Chalovich, L. E. Greene, and E. Eisenberg, Proc Natl Acad Sci USA **79**, 7288 (1982).

[62] K. Takiguchi, H. Hayashi, E. Kurimoto, and S. Higashi-Fujime, J Biochem **107**, 671 (1990).

[63] A. H. Krall and D. A. Weitz, Phys Rev Lett **80**, 778 (1998).

[64]    L. Cipelletti, L. Ramos, S. Manley, E. Pitard, D. A. Weitz, E. E. Pashkovski, and M. Johansson, Faraday Discuss **123**, 237 (2003).

[65]    D. A. Sessoms, H. Bissig, A. Duri, L. Cipelletti, and V. Trappe, Soft Matter **6**, 3030 (2010).

[66]    M. Soares e Silva, B. Stuhrmann, T. Betz, and G. H. Koenderink, New J Phys **16**, 075010 (2014).

[67]    E. Vincent, J. Hammann, M. Ocio, J.-P. Bouchaud, and L. F. Cugliandolo, in *Complex Behavior of Glassy Systems*, edited by M. Rubi (Springer Verlag, Berlin, 1997).

[68]    L. Cipelletti and L. Ramos, J Phys Condens Matter **17**, R253 (2005).

[69]    A. Duri, H. Bissig, V. Trappe, and L. Cipelletti, Phys Rev E **72**, 051401 (2005).

[70]    L. Cipelletti, H. Bissig, V. Trappe, P. Ballesta, and S. Mazoyer, J Phys Condens Matter **15**, S257 (2002).

[71]    A. Duri and L. Cipelletti, Europhys Lett **76**, 972 (2006).

[72]    M. Bouzid, J. Colombo, L. V. Barbosa, and E. Del Gado, Nature Commun **8**, 15846 (2017).

[73]    L. Berthier, G. Biroli, J.-P. Bouchaud, L. Cipelletti, and W. van Saarloos, editors, *Dynamical Heterogeneities in Glasses, Colloids, and Granular Media* (Oxford University Press, 2011).

[74]    T. T. Falzone, M. Lenz, D. R. Kovar, and M. L. Gardel, Nature Commun **3**, 861 (2012).

[75]    K. M. Schmoller, O. Lieleg, and A. R. Bausch, Soft Matter **4**, 2365 (2008).

[76]    D. Humphrey, C. Duggan, D. Saha, D. Smith, and J. Käs, Nature **416**, 413 (2002).

[77]    M. Mak, M. H. Zaman, R. D. Kamm, and T. Kim, Nature Commun **7**, 10323 (2016).

[78]    S. L. Freedman, G. M. Hocky, S. Banerjee, and A. R. Dinner, Phys Rev E **14**, 7740 (2018).

[79]    P. Lang and E. Frey, Nature Commun **9**, 494 (2018).

[80]    M. Sheinman, A. Sharma, J. Alvarado, G. H. Koenderink, and F. C. MacKintosh, Phys Rev E **92**, 012710 (2015).

[81]    Y. Mulla, G. Oliveri, J. T. B. Overvelde, and G. H. Koenderink, arXiv **cond-mat.soft**, 128 (2018).

[82]    M. Ohnaka, *The Physics of Rock Failure and Earthquakes* (Cambridge University Press, 2013).

[83]    I. Reiweger and J. Schweizer, Geophysical Research Letters **37**, L24501 (2010).

[84]    F. Gobeaux, E. Belamie, G. Mosser, P. Davidson, and S. Asnacios, Soft Matter **6**, 3769 (2010).

[85]    L. Cipelletti, S. Manley, R. C. Ball, and D. A. Weitz, Phys Rev Lett **84**, 2275 (2000).

[86]    G. Brambilla, S. Buzzaccaro, R. Piazza, L. Berthier, and L. Cipelletti, Phys Rev Lett **106**, 118302 (2011).

[87]    G. Nava, M. Rossi, S. Biffi, F. Sciortino, and T. Bellini, Phys Rev Lett **119**, 078002 (2017).

[88]    B. S. Gentry, S. Meulen, P. Noguera, B. Alonso-Latorre, J. Plastino, and G. H. Koenderink, Eur Biophys J **41**, 979 (2012).

[89]    A. Philippe, S. Aime, V. Roger, R. Jelinek, G. Prévot, L. Berthier, and L. Cipelletti, J Phys Condens Matter **28**, 075201 (2016).

[90]    B. J. Berne and R. Pecora, *Dynamic Light Scattering: with Applications to Chemistry, Biology, and Physics* (Dover Publications, 2013).

# Uncovering the dynamic precursors to motor-driven contraction of active gels

José Alvarado[1,2,$], Luca Cipelletti[3,*], Gijsje Koenderink[1,*]

[1] AMOLF, Physics of Living Matter Department, 1098 XG Amsterdam, Netherlands
[2] Massachusetts Institute of Technology, Department of Mechanical Engineering, Cambridge, MA, 02139, USA
[3] L2C, Univ. Montpellier, CNRS, Montpellier, France
[$] Current address: University of Texas at Austin, Department of Physics, Austin, TX, 78712, USA
[*] Corresponding authors. LC: luca.cipelletti@umontpellier.fr; GK: g.koenderink@amolf.nl

# SUPPLEMENTAL MATERIAL

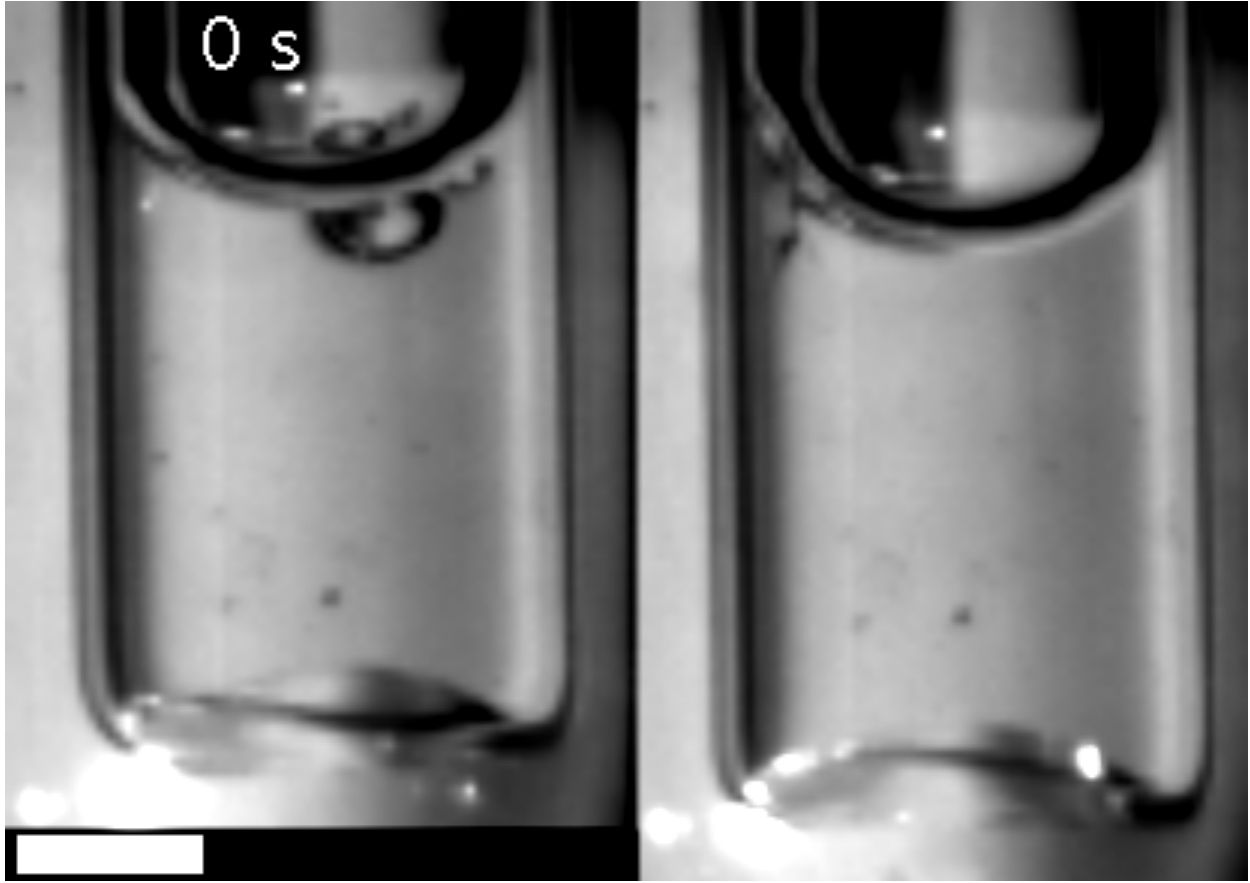

**Movie 1**. Two examples of contracting actin-myosin gels. The two samples have an identical biochemical composition ([fascin] = 0.6 μM, [myosin] = 0.06 μM, [actin] = 12 μM, [ATP] = 0.1 mM). Same samples as Main Text Figures 2, 3, and 4. (left) A free contraction event, where the gel adheres only to the top surface. (right) A pinned contraction event, where the gel initially adheres both to top and bottom surfaces. Scale bar 2 mm. Sample age given in top-left corner.

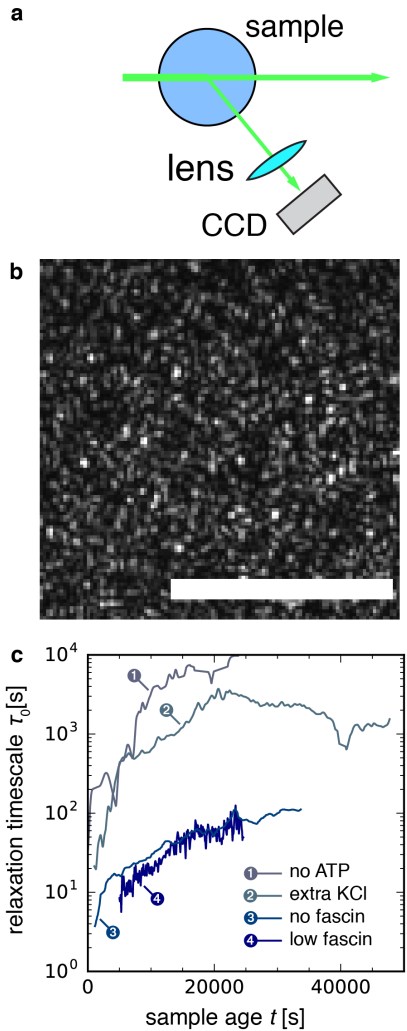

**Figure S1**. Relaxation timescale for non-contracting samples. (a) Schematic of light scattering setup, top view. Scattered light is acquired at angle of 45°, corresponding to scattering vector $q$ = 12.0 μm$^{-1}$. (b) Typical speckle pattern recorded by the CCD. Full images contain ca. 500–700 speckles. Scale bar 200 μm. (c) Temporal evolution of the microscopic relaxation time $\tau_0$ for one passive sample (labeled 1) and three active but non-contracting samples (labeled 2, 3, 4). Data shown for scattering vector $q$ = 12.0 μm$^{-1}$. The biochemical compositions of the samples were as follows: *1. No ATP*: [fascin] = 0.6 μM, [myosin] = 0.06 μM, [actin] = 12 μM); *2. Extra KCl*: [fascin] = 0.6 μM, [myosin] = 0.06 μM, [actin] = 12 μM, [KCl] = 150 mM; *3. No fascin*: [fascin] = 0 μM, [myosin] = 0.06 μM, [actin] = 12 μM; *4. Low fascin*: [fascin] = 0.24 μM, [myosin] = 0.06 μM, [actin] = 12 μM. Note that none of the samples contract and all of them exhibit only aging, as characterized by a monotonic increase of $\tau_0$ with $t$. The relaxation time has been obtained by fitting correlation functions averaged over a time interval $\Delta t$ = 50 s for samples 1–3, and 10 s for sample 4.

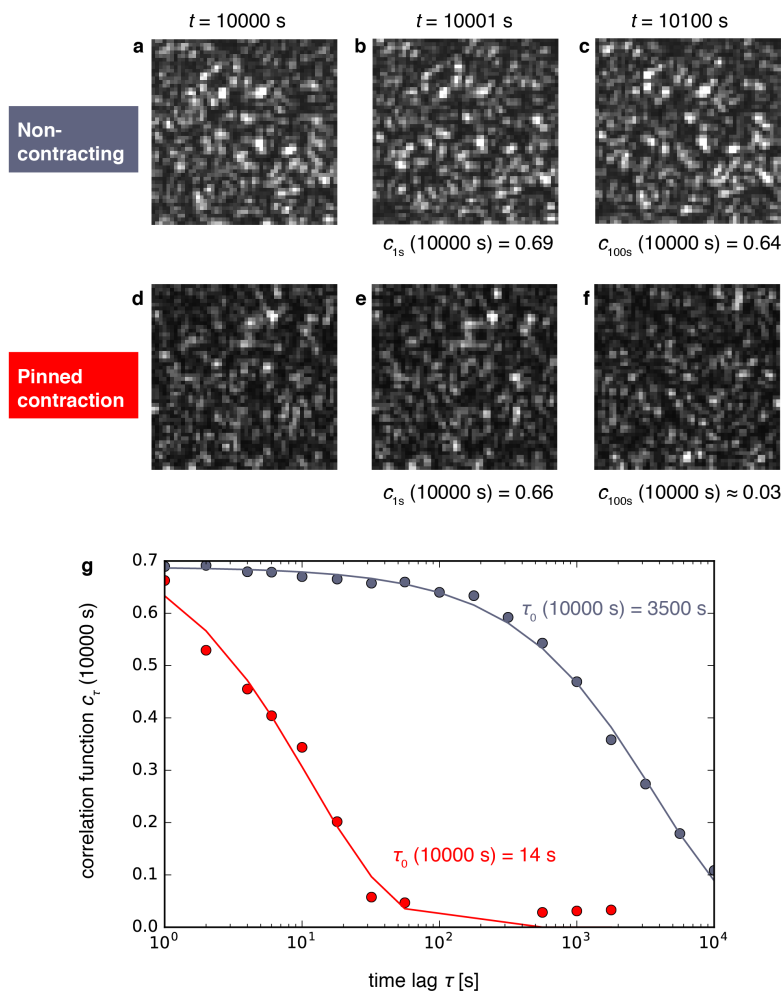

**Figure S2**. Overview of correlation analysis. (a–c) Snapshots of speckles of the non-contracting "No ATP" sample from Fig. **S1**. This sample is relatively static, which is reflected in the similarity of the speckle images taken at sample ages $t$ = 10000 s (a), $t$ = 10001 s (b), and $t$ = 10100 s (c). The degree of correlation $c_\tau$ (10000 s) with respect to image a) is displayed below images b) and c). Note that over the course of 100 s, the speckle image does not significantly change, and the intensity correlation function attains values of ~0.6. (d–f) Snapshots of speckles of the "pinned contraction" sample shown in the main text Figs. 2–5. This sample is more dynamic than the "No ATP" sample, which is reflected in the faster time evolution of the speckle images. Note the significant rearrangement of speckles in panel f compared to panels d and e, and the corresponding drop in the degree of correlation to ~0.03. (g) Determination of the relaxation timescale $\tau_0$. Plot of the intensity correlation function $c_\tau$ (at a sample age of 10000 s) as a function of the time lag $\tau$, for the "No ATP" (grey symbols) and the "pinned contraction" (red symbols) samples. Fits (lines) of data (circles) to a stretched exponential decay function yield the microscopic relaxation timescale $\tau_0$ that is plotted in the main text figures. Stretching exponent for "NoATP" sample: 0.72±0.03; for "pinned contraction" sample: 0.71±0.1

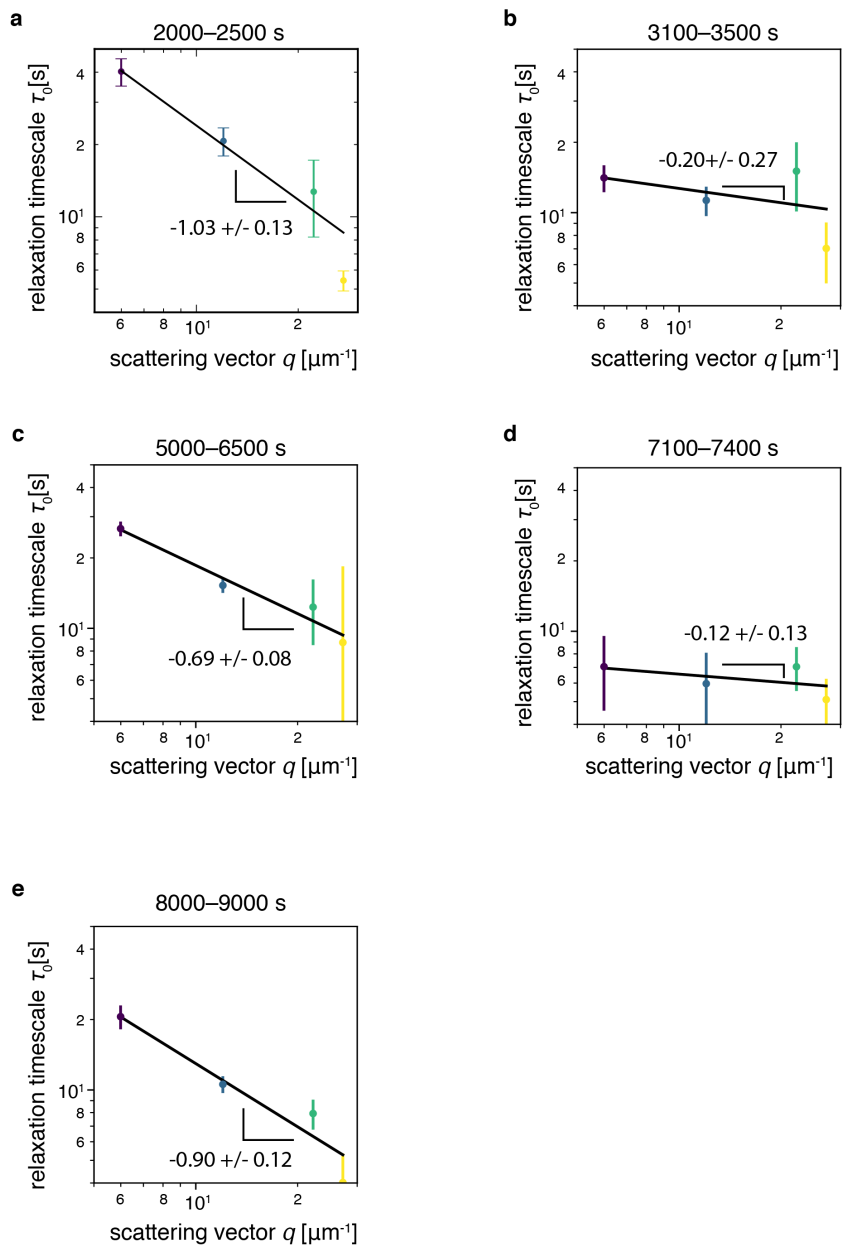

**Figure S3.** Plot of the relaxation timescale $\tau_0$ as a function of the scattering vector $q$ over various sample age windows: (a) 2000–2500 s, (b) 3100–3500 s, (c) 5000–6500 s, (d) 7100–7400 s, and (e) 8000–9000 s. Same contractile active gel sample as in Fig. 1c of the main text.

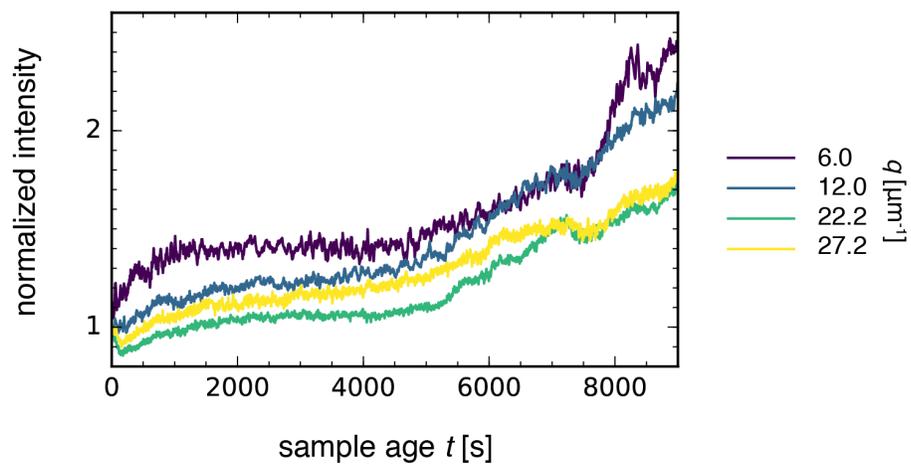

**Figure S4**. Normalized intensity $I / I_0$ as a function of sample age $t$. The four curves correspond to different scattering vectors $q$ (see legend, right). Same contractile active gel sample as in Fig. 1c of the main text.

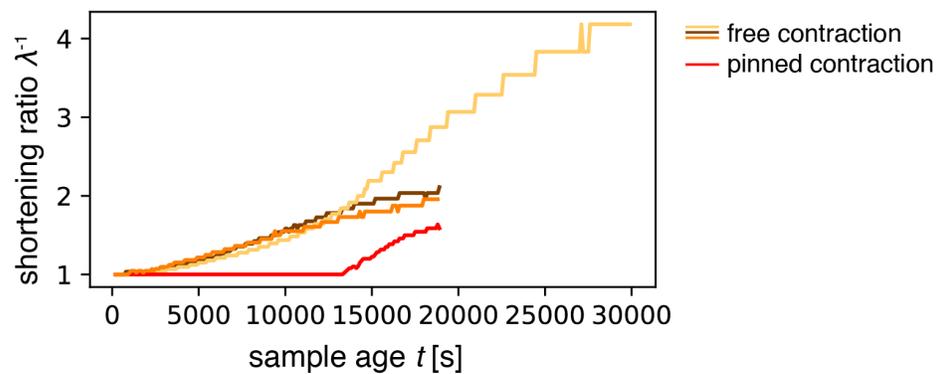

**Figure S5**. Contraction ratio $\lambda^{-1}$ as a function of sample age $t$ for free and pinned contraction events (see legend, right). Samples have [actin] = 12 µM, [ATP] = 0.1 mM, [fascin] = 0.6 µM, and [myosin] = 0.06 µM (orange, light orange, red curves) or [myosin] = 0.12 µM (brown curves).

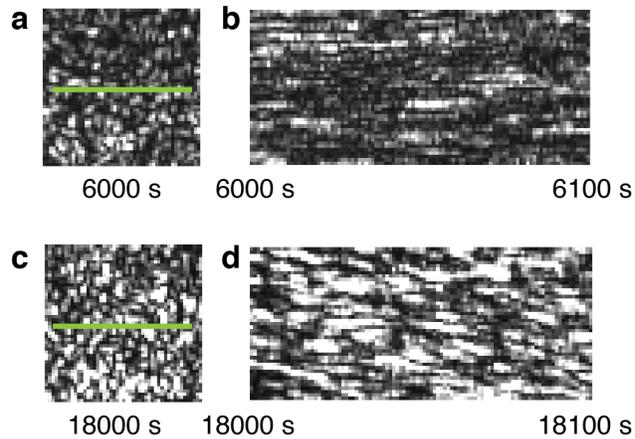

**Figure S6**. Demonstration of speckle drift for the free contracting sample depicted in Figure **3**. (a) Speckle pattern at 6000 s. (b) Kymograph of panel a, green line. Left column of pixels corresponds to a sample age of 6000 s, before contraction. Right column, 6100 s. Note the horizontal orientation of streaks, which indicates no spatial drift of speckles. (c) Speckle pattern at a sample age of 18000 s, during contraction. (d) Kymograph of panel c, green line. Left column of pixels corresponds to sample age 18000 s. Right column, 18100 s. Note the diagonal orientation of streaks, which indicates spatial drift of speckles.

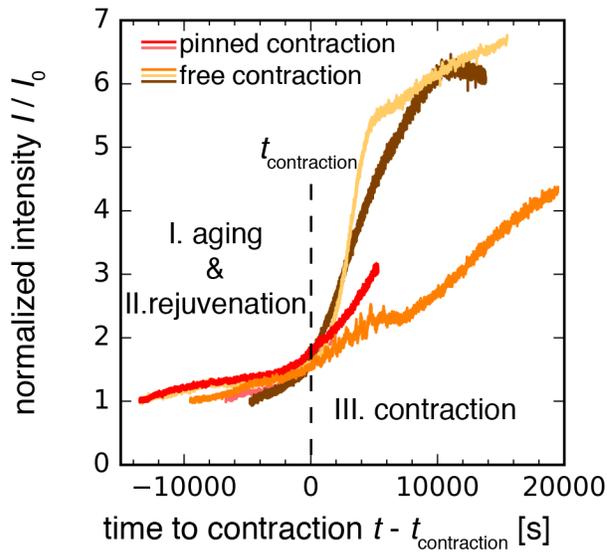

**Figure S7**. Average scattered intensity $I$ normalized by the initial average intensity $I_0$ as a function of sample age $t$ for all pinned contraction (red curves) and free contraction (orange curves) samples investigated (see legend, top left). Samples have [actin] = 12 μM, [ATP] = 0.1 mM, [fascin] = 0.6 μM, and [myosin] = 0.06 μM (orange, light orange, red curves) or [myosin] = 0.12 μM (brown and pink curves). The vertical dashed line indicates the onset of contraction as defined by the normalized intensity attaining a value of 1.5, which separates the stages of aging (I) and rejuvenation (II) from the contraction stage (III).

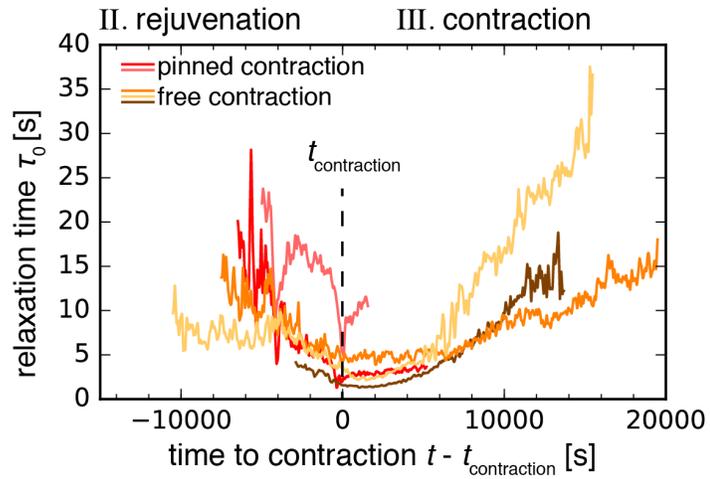

**Figure S8**. Relaxation time $\tau_0$ as a function of time to contraction, $t - t_{contraction}$, for the rejuvenation and contraction stages. Dashed vertical line denotes $t_{contraction}$, which separates stage II from stage III. Note that the beginning of the rejuvenation stage can occur up to 10000 s before contraction. Samples are composed of [actin] = 12 μM, [ATP] = 0.1 mM, [fascin] = 0.6 μM, and [myosin] = 0.06 μM (orange, light orange, red curves) or [myosin] = 0.12 μM (brown and pink curves).

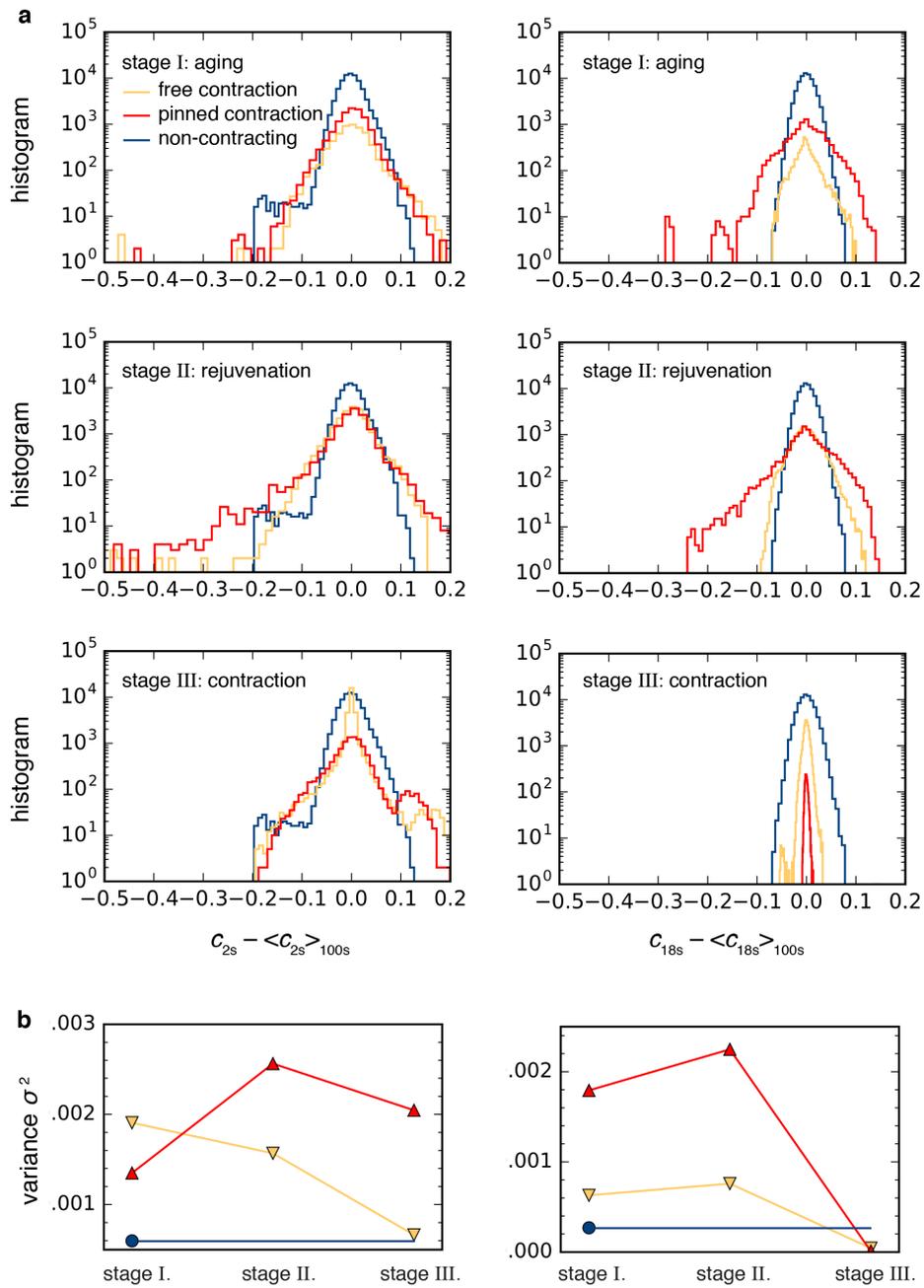

**Figure S9**. Statistics of the degree of correlation fluctuations $c_\tau - \langle c_\tau \rangle_{100s}$ for the three stages of sample evolution. (a) Histogram of $c_\tau - \langle c_\tau \rangle_{100s}$ for two different lag times, $\tau = 2s$ (left column) and $\tau = 18s$ (right column), broken down by stage. (Cf. main text, Fig. **5**d, where $\tau = 6s$.) (b) Corresponding variance $\sigma^2$ for lag times $\tau = 2s$ (left) and $\tau = 8s$ (right). Color and symbol denote sample type: non-contracting (blue circle), free contraction (orange downward triangles), and pinned contraction (red upward triangles).

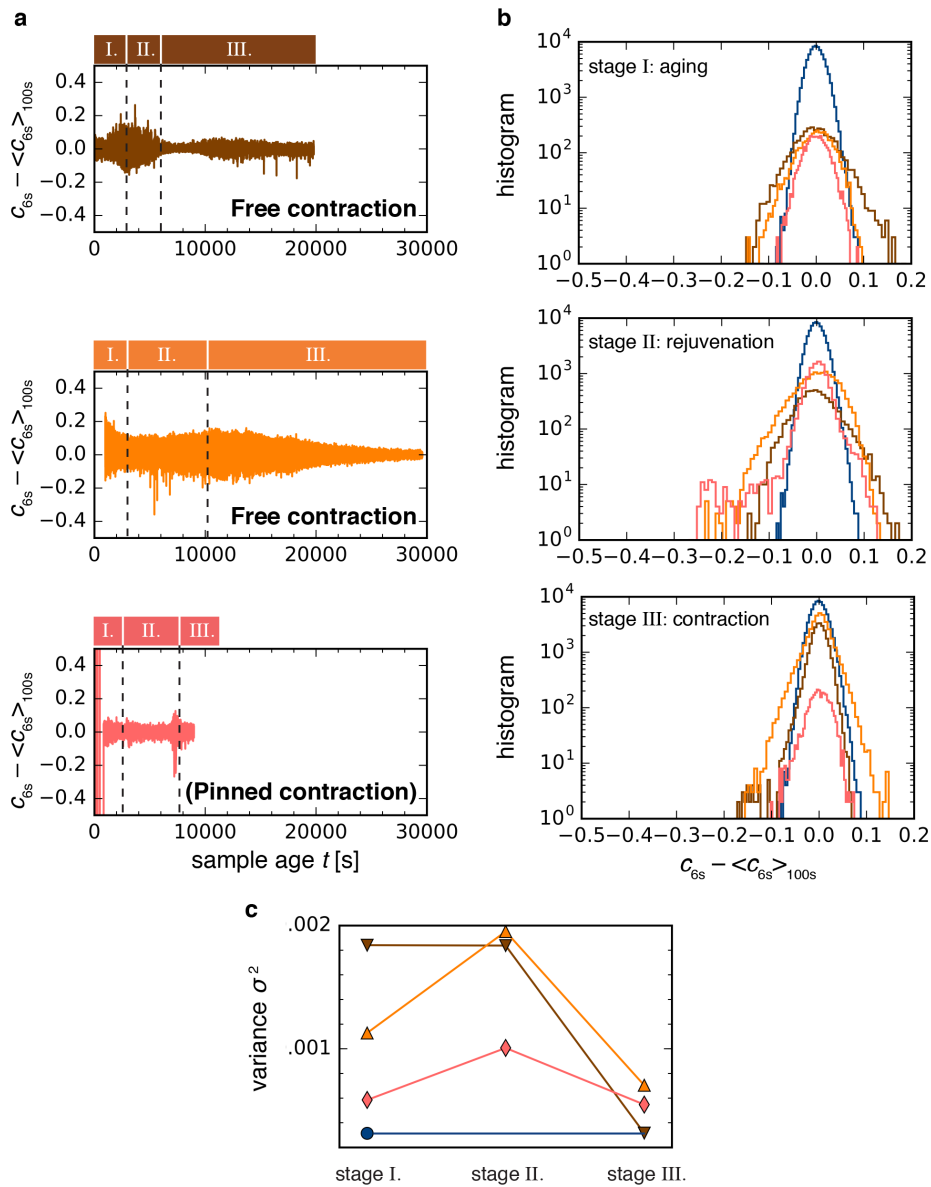

**Figure S10**. Statistics of samples not covered in the main text. The increased variance in stage II depicted in Fig. **5**e also holds for these samples. (a) Degree of correlation, as in Fig. **5**a–c. (b) Histograms for the three samples and the three stages shown in panel a. For reference, the same non-contracting sample from Fig. **5** is also shown here. (c) Variance $\sigma^2$ of the distributions shown in panel (b): free contraction (brown downward triangles), free contraction (orange upward triangles), contraction from Fig. **1** (pink diamonds), and the non-contracting sample from Fig. **5** (blue circle).